\documentclass[aps, prl, twocolumn, nofootinbib, nobibnotes]{revtex4-2}

\usepackage{amsmath}
\usepackage{amsfonts}
\usepackage{wasysym}
\usepackage{graphicx}
\usepackage{comment}
\usepackage{tensor}
\usepackage[svgnames]{xcolor}

\usepackage[colorlinks = true, linkcolor = DarkBlue, citecolor = DarkBlue, urlcolor = DarkBlue, bookmarks = false]{hyperref}

\def \EREPR {$\text{ER}=\text{EPR}$}

\def\filetype{pdf}
\def\path{}

\allowdisplaybreaks[1]

\begin{document}


\title{Probing the connection between entangled particles\\ and wormholes in general relativity}
\author{Ben Kain}
\email{bkain@holycross.edu}
\affiliation{Department of Physics, College of the Holy Cross, Worcester, Massachusetts 01610, USA}

\begin{abstract}
\noindent Maldacena and Susskind conjectured that two entangled particles, which can be thought of as forming an Einstein-Podolsky-Rosen (EPR) pair, are connected by a nontraversable wormhole or Einstein-Rosen (ER) bridge.  They named their conjecture \EREPR.  We present a concrete quantitative model for \EREPR, in which two spin-1/2 particles in a singlet state are connected by a nontraversable wormhole in asymptotically flat general relativity.  In our model, the fermions are described by the charged Dirac equation minimally coupled to gravity.  This system has static wormhole solutions.  We use these solutions as initial data and numerically evolve them forward in time.  Our simulations show that black holes form, which are connected by the wormhole and which render the wormhole nontraversable.  We also find that the wormhole throat shrinks, which places the particles in close proximity to one another and suggests an explanation for how the wormhole facilitates the nonlocal communication required by entanglement.
\end{abstract} 

\maketitle


\textit{Introduction}.\textbf{---}Maldacena and Susskind conjectured that entanglement, a cornerstone of quantum mechanics, has a gravitational explanation \cite{Maldacena:2013xja}.  They conjectured that two entangled objects, be it two entangled black holes or two entangled particles, which can be thought of as forming an Einstein-Podolsky-Rosen (EPR) pair, are connected by a wormhole or Einstein-Rosen (ER) bridge.  They named their conjecture \EREPR.

Entanglement requires some form of nonlocal communication, although this nonlocality cannot be used to send messages faster than the speed of light.  \EREPR\ posits that the nonlocal communication occurs through the wormhole.  The wormhole is therefore expected to be nontraversable, otherwise messages could be sent through the wormhole, and hence via entanglement, faster than light, violating the properties of entanglement.

If \EREPR\ is true, it revolutionizes our understanding of quantum mechanics and makes an extraordinary connection between quantum mechanics and gravity.  Much of the work on \EREPR\ has been on black holes in AdS, where AdS/CFT \cite{Maldacena:1997re} can be used (for a sampling of references on \EREPR, see \cite{Gharibyan:2013aha, Chernicoff:2013iga, Susskind:2014yaa, Gao:2016bin, Chen:2016nvj, Maldacena:2017axo, Susskind:2017nto, Maldacena:2018gjk, Anderson:2020vwi, Jafferis:2021ywg, Balasubramanian:2020hfs, Dai:2020ffw, Jafferis:2022crx}).  Entanglement, of course, is ubiquitous in quantum mechanics, occurring in a wide variety of systems.  Arguably the simplest example of entanglement is two spin-1/2 particles in a singlet state.

We present a concrete quantitative model for \EREPR\ in asymptotically flat general relativity.  In our model, two charged spin-1/2 particles in a singlet state are described by the charged Dirac equation minimally coupled to gravity.  This system has static wormhole solutions \cite{Blazquez-Salcedo:2020czn, Konoplya:2021hsm}.  The asymmetric solutions are smooth, regular everywhere, and violate the null energy condition, which suggests that they are traversable.  We use the asymmetric solutions as initial data and numerically evolve them forward in time.  Our dynamical solutions show  that black holes form, which are connected by the wormhole and which render the wormhole nontraversable.  Our model is therefore a system in which two spin-1/2 particles in a singlet state are connected by a nontraversable wormhole in asymptotically flat general relativity.

Maldacena and Susskind conjectured that a wormhole connecting two entangled particles would be ``very quantum" \cite{Maldacena:2013xja}.  Remarkably, the wormhole geometry in our model has a completely classical description.  The particles are described at the level of quantum wave functions, where we impose the one particle condition, consistent with the Pauli exclusion principle \cite{Finster:1998ws, Herdeiro:2017fhv, Blazquez-Salcedo:2020czn, Blazquez-Salcedo:2021udn}.  Under this condition, we show results for particles with a mass a few orders of magnitude smaller than the Planck mass and a wormhole throat radius a few orders of magnitude larger than the Planck length.

Alice and Bob could each be sitting just outside a mouth of the wormhole.  Alice and Bob cannot send messages to one another through the wormhole, as long as they stay outside, since the wormhole is nontraversable.  The wormhole is too small for human travelers, but, figuratively, it is possible for Alice and Bob to jump into the wormhole from opposite ends and meet each other inside.  We confirm these conclusions in our model by computing null geodesics.  Geodesics that travel through the wormhole become trapped inside a black hole, consistent with Alice and Bob being unable to send messages to one another while staying outside.  There exist geodesics that travel into the wormhole from opposite sides and cross, analogous to Alice and Bob meeting each other inside.

Our simulations show that a portion of the wormhole throat shrinks.  This shrinking places the two particles, which are on opposite ends of the wormhole, right next to each other.  It is tempting to think that this close proximity of the particles is how the wormhole facilitates the nonlocal communication required by entanglement.


\textit{Model}.\textbf{---}We consider two particles, both with mass $\mu$ and charge $e$, in a singlet state
\begin{equation}
\Psi = \psi_+ \wedge \psi_-
= \psi_+ \otimes \psi_- - \psi_- \otimes \psi_+,
\end{equation}
where $\psi_\pm$ are single-particle Dirac spinors.  The ansatz we use for $\psi_\pm$ will be given shortly and leads to the following Lagrangian for our matter sector \cite{Finster:1998ju},
\begin{equation} \label{Lagrangian}
\mathcal{L} = \sum_{x=\pm} \mathcal{L}_\psi^x + \mathcal{L}_\mathcal{A},
\end{equation}
where
\begin{equation} \label{Lagrangian 2}
\begin{split}
\mathcal{L}_\psi^x &= \frac{1}{2} \left[
\bar{\psi}_x \gamma^\mu D_\mu \psi_x
- (D_\mu \bar{\psi}_x) \gamma^\mu \psi_x
\right]
- \mu \bar{\psi}_x \psi_x
\\
\mathcal{L}_\mathcal{A} &= -\frac{1}{4} F_{\mu\nu} F^{\mu\nu}
\\
F_{\mu\nu} &= \partial_\mu \mathcal{A}_\nu - \partial_\nu \mathcal{A}_\mu,
\end{split}
\end{equation}
with $\mathcal{A}_\mu$ the $U(1)$ gauge field.  As shown in \cite{Blazquez-Salcedo:2020czn, Konoplya:2021hsm}, static wormhole solutions cannot be found without the presence of the gauge field.  We minimally couple the Lagrangian in Eq.\ (\ref{Lagrangian}) to gravity via $\mathcal{L} \rightarrow \sqrt{-\det(g_{\mu\nu})} \, \mathcal{L}$, where $g_{\mu\nu}$ is the metric.  We ignore second quantization effects and treat both the gauge field and gravity classically.  This has obvious drawbacks, but it also has some important advantages.  In particular, it allows us to straightforwardly take into account the back reaction of gravity on spinors.

We assume a standard form for the spherically symmetric metric,
\begin{equation} \label{metric}
ds^2 = -\alpha^2 dt^2 + A dr^2 + C \left( d\theta^2 + \sin^2\theta \, d\phi^2 \right),
\end{equation}
where $\alpha(t,r)$, $A(t,r)$, and $C(t,r)$ are functions of $t$ and $r$ and, for a wormhole geometry, $-\infty < r < \infty$.  The areal radius is given by $R = \sqrt{C}$ and we take $R(t,0)$, which gives the minimum value of $R$ on a time slice, to be the wormhole throat radius.

The Dirac spinor ansatz we use for $\psi_\pm$ is
\begin{equation} \label{Dirac ansatz}
\psi_\pm = \frac{e^{\pm i\phi/2}}{2\sqrt{\pi} A^{1/4}(t,r)C^{1/2}(t,r)}
\begin{pmatrix}
\hphantom{\pm i} F(t,r) y_\pm(\theta) \\
\pm iF(t,r) y_\mp(\theta) \\
\hphantom{\pm i}G(t,r) y_\pm (\theta) \\
\mp iG(t,r) y_\mp(\theta)
\end{pmatrix},
\end{equation}
where $y_+ \equiv \sin(\theta/2)$ and $y_- \equiv \cos(\theta/2)$.  The derivation of this ansatz is lengthy and we do not present it here.  We present it in Appendix B of \cite{kain_EDM}, where we list the precise set of assumptions we make, including our specific choice of spin-weighted spherical harmonics.  Fermions are parametrized in terms of the complex functions
\begin{equation}
\begin{split}
F(t,r) &= F_1(t,r) + i F_2(t,r), 
\\
G(t,r) &= G_1(t,r) + i G_2(t,r),
\end{split}
\end{equation}
where $F_{1,2}$ and $G_{1,2}$ are real.  

We simulate this system by solving the equations of motion for the Lagrangian in (\ref{Lagrangian}) and the Einstein field equations for the metric in (\ref{metric}).  The field equations depend on the energy-momentum tensor, which is computed from the Lagrangian in (\ref{Lagrangian}).  This constitutes a lengthy set of equations which are given in \cite{kain_EDM}, both for the general spherically symmetric metric and for the specific form in (\ref{metric}).  

We use the following dimensionless variables,
\begin{equation} \label{scaling}
\bar{r} \equiv \frac{r}{R_0}, \quad
\bar{t} \equiv \frac{t}{R_0}, \quad
\bar{e} \equiv \frac{R_0}{\sqrt{G}} e, \quad
\bar{\mu} \equiv R_0\mu, \quad
\overline{R} \equiv \frac{R}{R_0},
\end{equation}
along with $\overline{F}_{1,2} \equiv \sqrt{G/R_0} \, F_{1,2}$ and $\overline{G}_{1,2} \equiv \sqrt{G/R_0} \, G_{1,2}$, where $R_0 \equiv \sqrt{C(0,0)}$ is the initial wormhole throat radius, $G = \ell_P^2$ is the gravitational constant, and $\ell_P$ is the Planck length.

As mentioned, we describe fermions using quantum wave functions \cite{Finster:1998ws, Herdeiro:2017fhv, Blazquez-Salcedo:2020czn, Blazquez-Salcedo:2021udn}.  Specifically, we require there to be one fermion of each type, $N_\pm = 1$, consistent with the Pauli exclusion principle.  Our scaling in (\ref{scaling}) leads to $\overline{N}_\pm = (\ell_P/R_0)^2 N_\pm$, where
\begin{equation}
\overline{N}_\pm  = \int_{-\infty}^\infty d\bar{r} \, \overline{\mathcal{N}},
\qquad
\overline{\mathcal{N}} = \overline{F}_1^2 + \overline{F}_2^2 + \overline{G}_1^2 + \overline{G}_2^2
\end{equation}
is straightforward to compute from a numerical solution.  Setting $N_\pm = 1$ gives
\begin{equation} \label{R0 eq}
R_0 = \frac{\ell_P}{\sqrt{\overline{N}_\pm}}.
\end{equation}
This equation allows us to compute the physical radius of the initial wormhole throat, $R_0$, by computing $\overline{N}_\pm$.

We use asymptotically flat static wormhole solutions as initial data for our simulations.  To solve for static solutions, we take the static limit of our equations.  We then make the coordinate choice $A(r) = 1$, so that our static metric is 
\begin{equation}
ds^2 = -\alpha^2(r) dt^2 + dr^2 + C(r) (d\theta^2 + \sin^2\theta d\phi^2).
\end{equation}
We present the general set of static equations as well as the set for our particular coordinate choice in \cite{kain_EDM}.  As expected, the null energy condition can be shown to be violated for these static solutions \cite{Blazquez-Salcedo:2020czn, kain_EDM}.  Our static solutions are parameterized in terms of the three constants $\bar{\mu}$, $\bar{e}$, and $f_0 \equiv \overline{F}_1(0,0)$.

The code we use to simulate our model is based on the code used in \cite{kain}.
Additional aspects of our model and code, including boundary conditions, our choice for time slicing, and numerical methods are given in \cite{kain_EDM, kain}.  


\begin{figure*}
\centering
\includegraphics[width=7in]{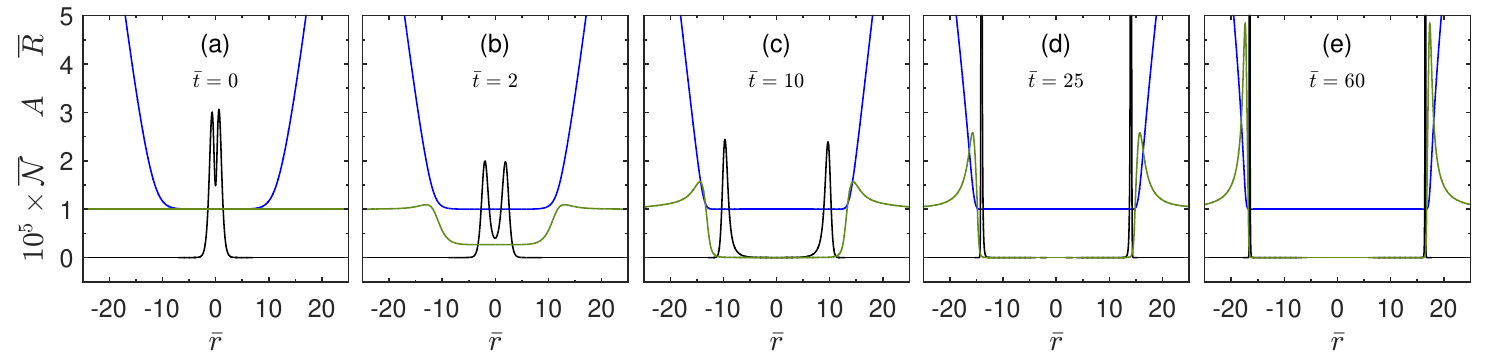}
\caption{This figure displays a few quantities for a typical simulation.  The black curve plots the number density, $\overline{\mathcal{N}}$.  The blue curve plots the areal radius, $\overline{R}$.  The green curve plots the metric field $A$.}
\label{fig:simulation}
\end{figure*} 

\begin{figure*}
\centering
\includegraphics[width=7in]{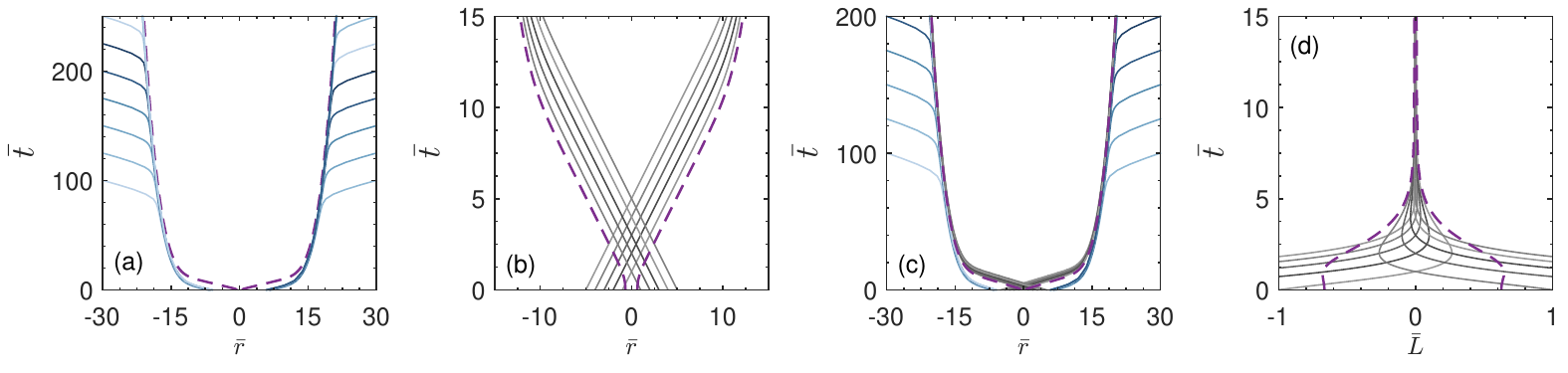}
\caption{An alternative view of the same simulation shown in Fig.\ \ref{fig:simulation}.  (a) The blue curves are null geodesics which map out event horizons for black holes on both sides of the wormhole.  The dashed purple curves plot the peaks of the number density shown in Fig.\ \ref{fig:simulation} and can be thought of as plotting the rough position of the two particles.  (b) The gray lines are null geodesics which travel through the wormhole.  (c) plots the same thing as (b), but over a larger range.  (d) plots the same thing as (b), but in terms of the physical distance from the origin $\bar{L}$ instead of in terms of the radial coordinate $\bar{r}$.}
\label{fig:null}
\end{figure*} 

\textit{Simulation.}\textbf{---}We show results for a typical simulation in Fig.\ \ref{fig:simulation}.  Figure \ref{fig:simulation}(a) displays a static wormhole solution defined by $\bar{\mu} = 0.2$, $\bar{e} = 0.1$, and $f_0 = 0.001$.  This solution has $R_0 = 375.3 \ell_P$ for the wormhole throat radius and $\mu = 0.00053 m_P$ for the particle mass, where $m_P$ is the Planck mass.

The black curve in Fig.\ \ref{fig:simulation} plots the (scaled) number density, $\overline{\mathcal{N}}$.  The number density can be interpreted as a particle distribution which means that we can interpret the two peaks in the black curve as representing two quasilocalized particles.  As time increases, the two particles separate with respect to their radial coordinates.  

The blue curve in Fig.\ \ref{fig:simulation} plots the areal radius, $\overline{R}$.  The areal radius gives a simple illustration of the wormhole on each time slice.  The horizontal portion of $\overline{R}$ depicts the length of the wormhole and the value of $\overline{R}$ at $\bar{r}=0$ gives the wormhole throat radius.  Caution should be taken with these interpretations, since they depend on the choice for time slicing.  We can see that the length of the wormhole increases, with respect to its radial coordinates, while the radius of the throat stays constant.  Our scaling convention sets $\overline{R}(0,0) = 1$.  If a null geodesic crosses $\bar{r} = 0$, our convention will be to say that it has traveled through the wormhole.

The green curve in Fig.\ \ref{fig:simulation} plots the metric field $A$.  Here we find particularly interesting behavior.  $A$ drops to zero along the wormhole and forms spikes at the mouths of the wormhole.  Since $A$ gives a measure of the physical distance between radial coordinates, this indicates that the physical length of a portion of the wormhole shrinks.  In fact, it appears that the physical distance between the two particles is the distance that is shrinking.  It is tempting to think that this shrinking distance, which puts the two particles in close proximity, is how the wormhole facilitates the nonlocal communication required by entanglement.

The wormhole is nontraversable, which requires the physical length of the wormhole to also increase \cite{Maldacena:2013xja}.  We can see this occurring with the spikes forming in $A$.  Such spikes are known as grid or slice stretching \cite{AlcubierreBook} and they are a standard indicator for the formation of a black hole.  The spikes in $A$ contribute to insurmountable physical distances, so that anything that travels through the wormhole is trapped inside a black hole and cannot escape.

Although the spikes in $A$ are a standard indicator for the presence of a black hole, we can make a conclusive determination for the presence or absence of a black hole.  In a numerical simulation, such a determination can be obtained by computing null geodesics backwards in time from the stored results of the simulation \cite{AlcubierreBook}.  If a black hole is present, these null geodesics will map out the event horizon.  We show such null geodesics in Fig.~\ref{fig:null}(a) as the blue lines, which clearly indicate the presence of black holes with event horizons on both sides of the wormhole.  We also include in Fig.~\ref{fig:null}(a) the location of the peaks of the number density in Fig.\ \ref{fig:simulation} as the dashed purple curves, which can be thought of as plotting the rough position of the particles.

In Figs.~\ref{fig:null}(b)--\ref{fig:null}(d), the dashed purple curves are again the location of the peaks of the number density in Fig.\ \ref{fig:simulation} and the gray lines are null geodesics we have computed in this geometry that travel through the wormhole.  In Fig.\ \ref{fig:null}(b), we can see null geodesics originating on both sides of the wormhole and traveling inward.  The geodesics are seen to travel through the wormhole and cross each other.  This result shows that Alice and Bob can (figuratively) jump into their respective mouths of the wormhole and meet each other inside.  Figure \ref{fig:null}(c) is the same as Fig.\ \ref{fig:null}(b), but plotted over a larger range.  We can see that the null geodesics are unable to travel arbitrarily far because they are trapped inside black holes.  This result illustrates how Alice and Bob cannot send messages to one another through the wormhole if they stay outside.

The (scaled) physical distance from the origin to radial coordinate $\bar{r}$ on a time slice is given by
\begin{equation}
\bar{L}(\bar{t}, \bar{r}) = \int_0^{\bar{r}} dx \, \sqrt{A(\bar{t},x)}.
\end{equation}
Figure \ref{fig:null}(d) is the same as Fig.\ \ref{fig:null}(b), but displays the curves in terms of their physical distance from the origin, $\overline{L}$.  We can see how the shrinking length of the wormhole affects the paths of the particles and geodesics.  Computing the physical distance $\bar{L}$ for the blue null geodesics in Fig.~\ref{fig:null}(a) is technically challenging and for this reason we do not include them in Fig.~\ref{fig:null}(d).


\textit{Discussion}.\textbf{---}Maldacena and Susskind's \EREPR~conjecture offers a radical rethinking of entanglement in quantum mechanics.  In \EREPR, entangled systems are connected by a wormhole and the nonlocal communication required by entanglement occurs through the wormhole.  This conjecture, as presented in \cite{Maldacena:2013xja}, was motivated in part by the fact that such a system offers explanations for various issues and paradoxes in quantum mechanics.  In this work, we presented a concrete model for \EREPR, in which two spin-1/2 particles in a singlet state are connected by a nontraversable wormhole in asymptotically flat general relativity.  In particular, the wormhole solutions followed directly from the Einstein field equations.

In our model, the two particles sit at opposite ends of the wormhole.  As the system evolves in time, the physical length of a portion of the wormhole throat shrinks such that the two particles are in close proximity to one other.  It is tempting to think that this is how the wormhole facilitates the nonlocal communication required by entanglement.

We computed null geodesics and showed that geodesics that travel across the wormhole are trapped inside a black hole.  This is analogous to Alice and Bob being unable to send messages to one another through the wormhole as long as they stay outside.  On the other hand, we found null geodesics that travel through the wormhole from opposite sides and which cross one another.  This is analogous to Alice and Bob jumping into the wormhole from opposite sides and meeting each other inside.

Our model for \EREPR\ has obvious inadequacies.  In particular, we treat gravity classically and ignore second quantization effects.  It would be both interesting and valuable to know the effect quantization has on these results \cite{Kain:2023jgu, Kain:2023pvp}.  Unfortunately, quantization with a time dependent metric used in a numerical simulation is challenging.  At the same time, it is surprising that there exists a classical description of a nontraversable wormhole connecting two particles in a singlet state.  We have also not studied the passage of matter through the wormhole.  Null geodesics are relatively simple to compute, which is the reason we considered them, but they do not take into account the effect a message sent through the wormhole would have on the wormhole itself.

An interesting goal for this model is to give additional insight into how the wormhole allows for the nonlocal communication required by entanglement.  Presumably this requires a continuous and smooth mechanism for measurement and wave function collapse that can be implemented in a numerical evolution.  Such a mechanism is beyond the scope of this Letter.  Nonetheless, we expect our model can act as a starting point for studying such phenomena.

We end this Letter with a comment about determining if a wormhole is traversable.  Our results show that violation of the null energy condition is insufficient for such a determination.  Black holes may form sufficiently quickly such that null geodesics are unable to travel through the wormhole without being caught inside a black hole, as they do in our system.  A time dependent analysis may be necessary to fully determine if a wormhole is traversable.




%

\end{document}